# Plagiarism Detection in arXiv

Daria Sorokina[1], Johannes Gehrke[1], Simeon Warner[2], Paul Ginsparg[2,3]
[1]Department of Computer Science, [2]Computing and Information Science, [3]Department of Physics
Cornell University, Ithaca, NY, USA
{daria, johannes, simeon, ginsparg}@cs.cornell.edu

## Abstract

*We describe a large-scale application of methods for finding plagiarism in research document collections. The methods are applied to a collection of 284,834 documents collected by arXiv.org over a 14 year period, covering a few different research disciplines. The methodology efficiently detects a variety of problematic author behaviors, and heuristics are developed to reduce the number of false positives. The methods are also efficient enough to implement as a real-time submission screen for a collection many times larger.*

## 1. Introduction

Improper reuse of text in academic research articles has a long, but largely undocumented, history. With the advent of widespread dissemination of electronic documents via the internet, the problem might be expected to worsen, due to the ease of obtaining and incorporating text written by others. On the other hand, those determined to appropriate text have long done so regardless of technology, whereas the greater ease of detection in the electronic realm may dissuade the less-determined. Full-text electronic research document corpuses have grown substantially over the past decade, and permit a systematic assessment of these issues. To our knowledge, there has been no systematic assessment even as to whether there is a background level of "borrowed" text snippets in a typical corpus which is ordinarily accepted by the community, and some threshold above which verbatim reuse would be regarded as inappropriate. In the following, we analyze these questions by making use of the arXiv: a document corpus that has nearly 100% coverage of certain research areas over an extended time period.

The arXiv is an archive of physics, mathematics, computer science and quantitative biology articles [5, 13]. Since its creation in 1991, the arXiv has grown to over 375,000 articles (as of 7 July 2006), currently growing at a rate of more than 4000 new submissions each month.

The arXiv corpus is an excellent testbed for various analyses because the vast majority of articles are in formats from which the text content can be extracted. Over 95% of articles are supplied as TeX source (including LaTeX and other variants). These are automatically processed to produce versions for display, with PDF by far the preferred format at present. This process yields a set of PDF files amenable to text extraction. Experience has shown that taking the text from PDF is more practical than using the TeX source directly, because of the varied and complex macro substitutions possible within TeX; in general a complete TeX system is needed to interpret these.

There have been a small number of cases of plagiarism reported to arXiv administrators by readers, some quite egregious. The availability of efficient algorithms, as described here, means that it will be possible to automate the detection process, both to identify plagiarism in the existing corpus, and to provide real-time detection of plagiarized articles at submission time.

The remainder of this paper is structured as follows. We first outline in Section 2 our approach for finding pairs of documents that have similar passages of text. Then in Section 3 we show the results of an extensive experimental study that guided our parameter selection for the task-specific tuning of document similarity. Our results show that our algorithms are effective in finding plagiarism in the arXiv. We discuss related work in Section 4, and we conclude in Section 5.

We emphasize that the contribution of this article is not the methodology of detecting plagiarism; in fact, there has been much previous work in this area, as we discuss in Section 4. To the best of our knowledge, however, this article describes the first large-scale application of methods for finding plagiarism in document collections, and it opens up interesting directions for future research. Due to space restrictions, we had to omit some implementation details and several experimental results; they can be found in the full version of this paper [12].

## 2. Finding similar documents

We first discuss the problem of finding passages of text that are held in common by two or more documents. Our corpus is a collection of *documents*; each document has one or several *authors*. In our simple model, a document is made up of *sentences*, each consisting of a sequence of *words*.

Sentences provide a natural partitioning of text, though with the drawback that a single sentence is not limited in length. A plagiarizer could slightly modify a long sentence so that the new sentence is formally different, but would still be identified as plagiarism by a human reader. To address this problem, we introduce the notion of *similar* sentences as having overlapping consecutive parts of some fixed size. Our next two definitions capture this intuition.

**Definition 2.1** *Two sentences are $\nu$-**similar** if they contain the same sequence of $\nu$ consecutive words.*

**Definition 2.2** *Two documents $D$ and $D'$ are $(\nu, m)$-**similar** if there exist $m$ sentences $s_1, \ldots, s_m$ in $D$, and $m$ sentences $s'_1, \ldots, s'_m$ in $D'$, such that $\forall i \in \{1, \ldots, m\}$, $s_i$ is $\nu$-similar to at least one $s'_j$ for $j \in \{1, \ldots, m\}$, and vice versa for $s'_i$. Note that we do not require that these sentences are consecutive in either of the documents.*

Shared sentences between documents, however, do not always indicate plagiarism. Some common sentences are used by many authors. For example, the sentence "This paper is organized as follows." occurs in a large number of documents, and such common sentences should be excluded from our analysis. To take this into account, we refine our definitions as follows:

**Definition 2.3** *A sequence of words is $L$-**common** if it is shared by at least $L$ documents with non-overlapping authorship. We call a sequence of words $L$-uncommon if it is not $L$-common.*

**Definition 2.4** *A sentence is $(\mu, L)$-**common** if it contains an $L$-common sequence of $\mu$ words. We call a sentence $(\mu, L)$-uncommon if it is not $(\mu, L)$-common.*

**Definition 2.5** *Two documents are $(\mu, L; \nu, m)$-**similar** if both contain at least $m$ different $(\mu, L)$-uncommon sentences that are $\nu$-similar to sentences in the other document.*

In the remainder of this section, we describe our approach for finding all document pairs that are similar by our definitions. Our approach is based on winnowing, a technique that has been previously used for plagiarism detection in programming assignments [9].

**The winnowing algorithm.** The winnowing algorithm is an instance of document fingerprinting: a document is summarized by a small set of character sequences called *fingerprints* which can be efficiently used to find copies of parts of a document in a large document collection. Document comparison is then reduced to finding exact matches in the sets of fingerprints.

The winnowing algorithm [9] is one specific instance of a fingerprinting algorithm. For a given document, the algorithm first selects all contiguous subsequences of characters of length $k$, called $k$-grams.

The set of fingerprints associated to a document is reduced significantly by considering the document as a maximal set of $n - w + 1$ overlapping *windows* of length $w$, and retaining only enough hashes to ensure that each window contains at least one hash. If $w$ is sufficiently large compared to $k$, the number of fingerprints associated to a document of length $n$ is significantly smaller than $n - k + 1$. It is straightforwardly shown [9] that even this reduced set of fingerprints can still find copies of sufficiently long passages of text, depending on the settings of parameters $k$ and $w$: any match at least as long as the *guarantee threshold* $t = w + k - 1$ will be detected.

**Text winnowing.** Winnowing was developed for arbitrary digital documents. The winnowing algorithm can be adapted to our document collection by taking advantage of the background knowledge that the dataset consists of text documents that can be naturally segmented into words and sentences. This permits selecting a much smaller initial set of possible $k$-grams, and a restricted set of windows, which will both increase the speed of the algorithm and also permit working with a smaller but nonetheless effective set of fingerprints.

The differences between the text and original versions of winnowing are as follows:

- The minimum unit of text is a word instead of a character. This means that the fingerprint represents a sequence of $k$ consecutive words rather than $k$ consecutive characters, and that the size of the window is measured as $w$ words, not characters.

- Each fingerprint is a subsequence of a sentence. Each window, from which a fingerprint is chosen, is a subsequence of a sentence as well. If a sentence is shorter than $k$ words, it is ignored. If its length is greater than or equal to $k$, but less than $w$ words, one $k$-gram is chosen from this sentence.

The second refinement eliminates many windows and fingerprints that would have to be calculated in the original algorithm — those that cross sentence boundaries. It also permits a more reliable measure of text overlap than just the number of shared fingerprints. At one extreme, $X$ shared fingerprints could correspond to a consecutive piece of only $k + X - 1$ words, because fingerprints can overlap. At the other extreme, when no fingerprints overlap, there may be

$X$ shared disjoint sequences of $k$ tokens apiece. This is an inherent inconvenience of the original algorithm and is dealt with here by counting the size of overlap as the number of similar *sentences*, instead of the number of shared fingerprints.

We now describe how text winnowing can be used to find all pairs of documents similar by definitions 2.2 and 2.5. By setting $t = \nu$, we can guarantee that all $\nu$-similar sentences will be detected. Then we calculate the number of sentences containing matching fingerprints and choose all documents that share at least $m$ such sentences. The resulting set of pairs will contain all $(\nu, m)$-similar documents.

We can easily extend the approach to find all $(\mu, L; t, m)$-similar documents by taking $\mu = k$. After determining the number of files containing each $k$-gram in its fingerprint set, we identify a subset of all $L$-common $k$-grams, and therefore a subset of all $(k, L)$-common sentences. After eliminating them from the analysis, we are left with a superset of all $(k, L)$-uncommon sentences, forming a superset of all $(k, L; t, m)$-similar pairs of documents.

**Parameter setting.** The experiments for choosing reasonable parameters $k$ and $t$ of the winnowing algorithm were run on a subset of the data, selecting only 7200 articles from the "physics" subject area of arXiv. Parameters $m$ and $L$ were set to 4 based on intuition and computational restrictions.

The general framework of these experiments was the following: We set $m = 4$, looking for documents that share at least four sentences containing uncommon $k$-grams, and ran our algorithm many times with different values of $k$ and $t$. We then manually assessed the differences between the sets of results to determine whether those pairs should be included or not, i.e., whether they constituted actual plagiarism or were false positives. These observations were used to select parameter values that screened as many false positives as possible, without losing too many true positives. Manual assessment of these differences was very time consuming so we used only a small subset of documents. We have chosen $k = 7$ and $t = 12$, these values appear to be adequate for present purposes, but a more comprehensive assessment would be required to determine optimal values.

Given all of the above, together with the discussion in Section 2, our algorithm ensures that we discover all (7,4;12,4)-similar pairs of documents.

## 3. Experiments

In our experiments, we used a collection of arXiv articles from 1991 through early 2005. This dataset included 287,857 articles in PDF and TeX format. Out of those, 3023 were unusable due to conversion problems. The experiments were run on a single CPU (64 bit, Itanium 2, 1.3 GHz) with 64 GB RAM. After preprocessing, the final run of the text winnowing algorithm and output of results took 20 hours. In the first step, 204,828,778 7-grams were entered into the multiple entry hash table, on average 721 7-grams per document. A set of 440,224 pairs of documents, each sharing at least 4 potentially uncommon 7-grams, was identified, and then reduced to 330,306 pairs sharing at least 4 potentially interesting similar sentences. This is our superset of all pairs of (7,4;12,4)-similar documents. Out of them, 17,621 are pairs of documents created by different authors and are therefore plagiarism candidates.

### 3.1. Extracting interesting documents

After creating the superset of all (12,4;7,4)-similar document pairs, we then extracted the subset relevant to the task of plagiarism detection.

This task turned out to be nontrivial because there are many reasons why shared text in two documents by different authors might not turn out to be plagiarism. A number of additional rules had to be applied in order to extract a set of pairs containing only a few false positives (i.e., non-plagiarism cases). Indications of possible false positives are as follows: (1) An author of one document is a neighbor on the co-author graph to an author of the other document. People from the same research group can both reuse text from an earlier article, on which they were co-authors. (2) An author of one article appears in the text of references of the other. This is usually an indication of "mild plagiarism" — people who maliciously claim someone's work as their own normally do not cite their source. (3) An author of one article appears in the text of the other article. This can result from any of:

- not all authors are properly indicated in the metadata
- there is a direct citation from the other article
- the first document is a workshop proceedings and the second is from the workshop
- one of the actual authors is mentioned only in the acknowledgments

(4) One or both articles are produced by a collaboration. Although significant effort was made to solve the collaboration problem during the preprocessing step, collaborations remain a probable source of false positives. (5) Both previous conditions (3,4) hold: this is a very strong evidence of a false positive, and occurs for example when all actual authors are listed in the full text of the article but not in the metadata.

These rules can be too strong, however, and could result in elimination of true positives. The pairs flagged by either heuristic 1 or 5 are discarded as false positives, but those flagged only by any combination of 2,3,4 are retained in a secondary set of results. The primary list contains only those pairs with strong evidence of plagiarism, and the secondary contains mainly false positives, but some cases of interest.

| heuristic | affected cases | impact |
|---|---|---|
| 1. coauthor | 8934 | 50.7% |
| 2. referenced | 6590 | 37.4% |
| 3. mentioned | 13148 | 74.6% |
| 4. collaboration | 2973 | 16.9% |
| 5. ment. & coll. | 2116 | 12.0% |

**Table 1. Rules applied to similar documents**

The impact of each rule is shown in Table 1. The second column lists the number of affected cases and the third shows the percentage of the affected cases in our set of pairs of overlapping documents with non-overlapping authorship (17621 pairs).

### 3.2. Results

**Common $k$-grams.** We found the document corpus to contain 429,258 common 7-grams: consecutive sequences of 7 words shared by at least 4 documents written by different authors. Table 2 shows the ten most frequently occurring $k$-grams, those that appeared in the largest numbers of documents. The numbers in this table count the appearances of these $k$-grams as (winnowed) fingerprints, and therefore can underestimate the true total numbers of appearances. These common 7-grams are instantly recognizable to people familiar with research literature in the subject area, and typically fall into classes such as: describing the structure of the article; describing equations, figures, tables; parts of common descriptive research phrases; acknowledgment text; or institutional affiliations.

**Plagiarism detection results.** Applying the heuristics of Section 3.1 to screen for false positives or "mild" plagiarism left 677 pairs of documents with at least four sentences sharing uncommon 7-grams. Detailed manual analysis of the first 20 pairs from this list revealed that 16 of them constitute clear cases of plagiarism. The 4 false positives resulted from i) two articles by the same author with two non-standard transliterations of his name, ii) two proceedings of different workshops sharing an article written by the same author, iii) two articles quoting the same text from Einstein, and iv) two articles with significant overlap in references which had not been automatically separable from the actual document texts.

Of the first 16 true positives, at least 3 appeared to be serious plagiarism, in which an article was essentially a copy of another written by others, albeit with many small text modifications. Many of the others were cases of articles and theses with an introductory or related work section appropriated from other sources, without even referencing them.

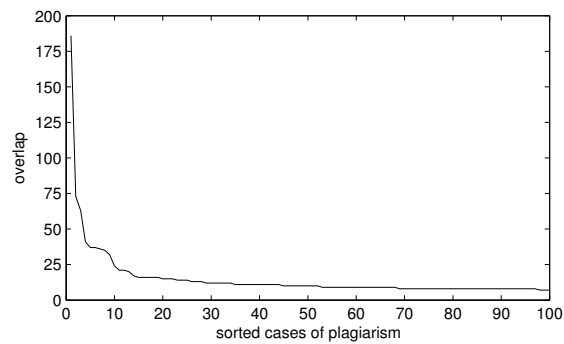

**Figure 1. Plagiarism cases sorted by overlap.**

Figure 1 shows the size of overlap (number of sentences sharing uncommon 7-grams) in the top 100 of the 677 flagged cases. Although there are relatively few cases with large chunks of text plagiarized, the number of cases becomes large as the size of the overlap decreases. The last 535 of the 677 cases came from 79 cases sharing 6 similar sentences, 138 sharing 5, and 318 sharing 4.

The 677 flagged pairs contain 1086 unique articles. Some of those appearing in multiple pairs are cases of a document plagiarizing from more than one source, and some from a source plagiarized multiple times: plagiarists often copy from many sources, and some sources, such as lecture notes, are attractive to many plagiarists.

There were an additional 7371 cases in the secondary list, removed from the primary list by the heuristics of Section 3.1. These are mainly false positives, but some examples of unethical behavior also appear. In the top 20 pairs, three were cases in which tens of pages were copied into a thesis without any modifications, although in these cases the sources were at least acknowledged in the text. (Proper usage, however, would require the entirety of the offending text to be placed in direct quotes.)

The secondary list contains 10,072 unique documents and the two lists together contain 10,763 unique documents.

**Visualization.** To identify interesting cases of plagiarism, it is useful to employ a graphical representation of the lists of document pairs. Many plagiarized articles contain text from different sources, and therefore appear in many entries of these lists. If the list is large, it is difficult to estimate the extent of plagiarism in a single article. To overcome this problem we have visualized the results as a graph of overlaps: each node is a document, edges connect overlapping documents, labels on edges show the number of shared similar sentences. Black edges correspond to the primary list of results (most probably plagiarism), grey edges correspond to the overlaps reported in the secondary list ("mild" plagiarism, if any at all). We have masked the ids of actual arXiv documents with single-letter labels: A, B, C, ... correspond to sources; Z, Y, X, W correspond to documents containing copied text. Which document in a pair contains original text and which contains its copy was decided manually based on submission dates and contents of papers.

| 7-gram | Documents | Authors | All occurrences |
|---|---|---|---|
| this work was supported in part by | 12966 | 2085 | 13161 |
| can be expressed in terms of the | 6541 | 2460 | 7379 |
| work was supported in part by the | 4612 | 1015 | 4760 |
| first term on the right hand side | 3337 | 1524 | 4000 |
| it is easy to see that is | 2974 | 1452 | 3605 |
| operated by the association of universities for | 2900 | 396 | 3372 |
| department of physics and astronomy university of | 2880 | 539 | 3018 |
| the paper is organized as follows in | 2764 | 1202 | 2766 |
| there is one to one correspondence between | 2418 | 1316 | 2967 |
| term on the right hand side of | 2404 | 1194 | 2846 |

**Table 2. Most popular 7-grams**

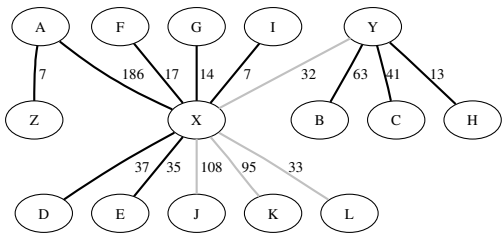

**Figure 2. Two major plagiarists.**

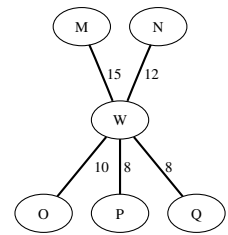

**Figure 3. A new case of real plagiarism.**

Figure 2 shows a subgraph corresponding to two particularly egregious cases of plagiarism. X is a PhD thesis, two thirds of which is copied from a variety of sources. Some sources are acknowledged, but most are not. Y is a journal article reviewing some area of physics. Several pieces of it are copied verbatim from other articles. It is an amusing coincidence that X and Y are connected: X copied in turn from Y.

Another node, A, has several adjacent edges for an opposite reason: A corresponds to lecture notes, and several people found it useful for their work. Apart from X which contains many pages stolen from A, another thesis, Z, shows a case of "mini-plagiarism": a small chunk of 7 sentences is copied from A.

The overlap graph was not really necessary to discover X and Y: they had copied sufficiently many pages, even from single sources, that they had appeared in the top cases in the list of pairwise overlaps. Figure 3, on the other hand, shows a less obvious case of plagiarism that is easily visualized in the graphical representation. Node W here corresponds to an article comparing different methods, and descriptions of the methods are copied from other articles. Each description is short, so the separate cases in the list of pairwise overlaps might not appear to be significant. Combined together, however, they indicate that a large part of the article's text is copied.

## 4. Related work

Due to space constraints, we can only give a very brief discussion of related work on plagiarism detection tools. Most previous work was either intended for and tested on small sets of documents (often less than 100), or was designed and used for different types of text, such as programming assignments. There has been less work on scalable systems tuned for large text document collections.

Schleimer, Wilkerson, and Aiken [9] invented the winnowing algorithm, a variation of which we use in this paper. Broder [2] used document fingerprints, but chose the smallest $k$-gram hashes from the entire document, not from smaller windows. Brin et al. [1] developed COPS, a system designed to detect copying in research articles. Koppel and Schler [6] suggested an approach to authorship detection based on gradually removing the most useful features of a text and comparing with other documents using only the remaining features. In CHECK [11], another plagiarism detection system for text documents, Si et al. address the problem of complexity of pairwise comparisons by in-troducing hierarchical comparison. Collberg et al.'s system SPLAT [3] crawls the websites of top CS departments and collects research articles as a dataset aimed at detecting self-plagiarism. Ribler and Abrams [8] suggested a method to visualize the degree of overlap of one document with a set of documents. The SCAM system [10] developed by Shivakumar and Garcia-Molina relies on word level analysis. Tuned to discover small overlaps, it results in many false positives when word distributions are similar but the texts are still different. The approach used in MatchDetectReveal system by Monostori et al. [7] avoids using a hash-function due to concern about hash collisions. In recent work, David and Pinch [4] used a modified version of our software to examine the extent and goals of copying and plagiarism in user reviews on amazon.com.

## 5. Conclusions

We identified over 500 cases of likely plagiarism from other authors, and additionally over 1000 cases of likely mild plagiarism. These constitute roughly 0.5% of the corpus, and an even smaller percentage of authors, since many come from repeat offenders. Some of the problems are quite serious, and many of the articles are published in conventional venues. None of the plagiarizers, the victims, nor their publishers have yet been notified, hence they remain anonymized here. We can, however, dispel some uncertainty by pointing out that while prominent (highly cited) authors are frequently victimized, they do not appear to reuse text from others.

The above results may tend to exaggerate the extent of the problematic behavior, but this needs further study. Many of the isolated copied sentences are of "background" nature, containing neither particularly unique information content nor stylistic virtue, and the "victims" might not even feel victimized. Some cases may reflect demographic and educational differences in an international author pool, with some careless reuse by non-native English writers who fear garbling content by modifying it. A next step underway is a private interface that displays the pairs of documents side by side, with overlapping text highlighted, and the ability to solicit confidential feedback from authors of both documents regarding the significance of the overlap. We can also envision a module for real-time screening of new submissions.

**Acknowledgements.** We thank Patrick Ng for his contributions to the data cleaning code. We also thank the Cornell Theory Center for usage of their equipment to run some of the experiments. This research was funded by NSF grants IIS-0636259, IIS-0541507, and by the KD-D Initiative. Any opinions, findings, conclusions or recommendations expressed in this material are those of the author(s) and do not necessarily reflect the views of the sponsors.